\documentstyle[rawfonts,nassau,pslatex]{article}  
\frompage{000} \topage{000}                                              
\def\rightmark{Thermal Lepton Production} 
\def\leftmark{Ralf Rapp}
 
\title{Thermal Lepton Production in Heavy-Ion Collisions} 
\authors{
{Ralf Rapp }\\[2.812mm]
{\normalsize
\hspace*{-8pt} Department of Physics and Astronomy,\\
SUNY Stony Brook, New York 11794-3800, U.S.A\\[0.2ex] 
}}
 
\abstract{The current status of evaluating thermal production
of lepton pairs in high-energy collisions of heavy nuclei is discussed. 
After a brief survey of emission rates from hot and dense matter, 
we address applications to (and interpretations of) recent SPS data 
including 40~AGeV CERES results, as well as prospects for NA60.  
Emphasis is put on predictions for RHIC. In particular, implications   
of hadronic observables and first single-electron measurements at 
$\sqrt{s}=130$~AGeV for upcoming pair spectra from PHENIX at 
$\sqrt{s}=200$~AGeV are assessed.} 

\keyword{Heavy-Ion Collisions, Dileptons, Chiral Symmetry Restoration} 
\PACS{25.75.-q, 12.38.Mh., 24.10.Pa}
 
\begin{document}
\newcommand{\beq}{\begin{equation}}
\newcommand{\eeq}{\end{equation}}
\newcommand{\bea}{\begin{eqnarray}}
\newcommand{\eea}{\end{eqnarray}}
\newcommand{\bfig}{\begin{figure}}
\newcommand{\efig}{\end{figure}}
\newcommand{\ie}{{\it i.e.}}
\newcommand{\bce}{\begin{center}}
\newcommand{\ece}{\end{center}}
\newcommand{\eg}{{\it e.g.}}
\newcommand{\Eg}{{\it E.g.}}
\newcommand{\etal}{{\it et al.}}
\def\lsim{\mathrel{\rlap{\lower4pt\hbox{\hskip1pt$\sim$}}
    \raise1pt\hbox{$<$}}}         
\def\gsim{\mathrel{\rlap{\lower4pt\hbox{\hskip1pt$\sim$}}
    \raise1pt\hbox{$>$}}}         
 
\maketitle
\setcounter{page}{1}

\section{Introduction}
\label{intro}
Early on in the heavy-ion initiative, electromagnetic observables have 
been recognized as important probes of the hot and dense strongly 
interacting systems created in head-on collisions of nuclei at high 
energies~\cite{Shu80,MT85}. About 10-15 years later very exciting data 
on dilepton (and photon) spectra have emerged from the SPS heavy-ion 
program~\cite{na45-158,na50-int,na45-40}, stimulating vigorous 
theoretical activity (see ref.~\cite{RW00} for a recent review). 
All respective experiments have observed nontrivial excess production 
in heavy-ion collisions (increasing with centrality) over baseline 
expectations from $p$-$p$ collisions (and/or final-state hadron decays),
from threshold to about 3~GeV (dilepton invariant mass or photon 
transverse momentum). Indeed, these results imply strong evidence 
for some of the long sought-for signatures, such as in-medium 
modified vector mesons (especially the $\rho$) at low-mass 
($M\le 1$~GeV) or continuum-like thermal radiation from early phases 
at intermediate masses (1~GeV~$\le M\le$~3~GeV).    
However, from the theoretical side, certain ambiguities in the 
detailed understanding and interpretation of the underlying dynamics 
remain at present. 
 
The article is organized as follows: in Sect.~\ref{sec_emcorr}
we briefly recall the theoretical framework suitable to evaluate  
equilibrium dilepton production rates and allude to specific 
model calculations.
In Sect.~\ref{sec_appl} we turn to applications in heavy-ion collisions, 
starting with a discussion on chemical off-equilibrium effects in the
hadronic phases in~\ref{sec_offeq}.  
In~\ref{sec_sps} we highlight results for SPS energies including 
predicted spectra for $E_{lab}=40$~AGeV as well as for a potential 
low-mass measurement by NA60.   
In~\ref{sec_rhic} we present improved predictions for RHIC
energies by incorporating information from measured hadron 
spectra. We also assess the role of thermal radiation in 
preliminary PHENIX single-electron spectra at $\sqrt{s}$=130~AGeV,
and ensuing implications for upcoming pair spectra.
In Sect.~\ref{sec_sum} we end with some concluding remarks.

\section{Electromagnetic Correlation Function and Dilepton Radiation}
\label{sec_emcorr}

\subsection{E.M. Correlator in Vacuum}
Dilepton emission from thermalized QCD matter can be characterized 
by the pair production rate per unit 4-volume and 4-momentum, 
\begin{equation}
\frac{dR_{l^+l^-}^{therm}}{d^4q} = -\frac{\alpha^2}{\pi^3 M^2}
 \ f^{B}(q_0;T) \ {\rm Im} \Pi_{\rm em}(M,q;\mu_B;T)
\label{rate}
\end{equation}
($f^B(q_0;T)=1/(\exp[q_0/T]-1)$: Bose distribution function, 
$M^2=q_0^2-q^2=(p_++p_-)^2$:  invariant mass squared of the lepton pair 
with individual 4-momenta $p_\pm$).
The key quantity encoding the information on the strongly interacting 
medium is the imaginary part of the electromagnetic current-current 
correlation function (or simply photon selfenergy), 
${\rm Im} \Pi_{em}$. In vacuum, it is directly related to the 
famous cross section ratio $R=\sigma_{ee\to hadrons}/\sigma_{ee\to\mu\mu}$
and can be decomposed into two invariant-mass regions:
\begin{equation}
{\rm Im} \Pi_{\rm em}(M) = \left\{
\begin{array}{ll}
 \sum\limits_{V=\rho,\omega,\phi} \left(\frac{m_V^2}{g_V}\right)^2 \
{\rm Im} D_V(M) & , \ M \le M_{dual}
\vspace{0.3cm}
\\
-\frac{M^2}{12\pi} \ (1+\frac{\alpha_s(M^2)}{\pi} +\dots)  \ N_c
\sum\limits_{q=u,d,s} (e_q)^2  & , \ M \ge M_{dual} \ .
\end{array}  \right.
\label{Piem}
\end{equation}
At small distances, \ie, $M\ge M_{dual}\simeq 1.5$~GeV, 
perturbation theory applies and the 
correlator strength follows from the 
(charge-square weighted) sum over free $q\bar q$ states (with rather small 
corrections); subsequent hadronization does not affect the strength. The 
experimental fact that hadron production in $e^+e^-$ annihilation
essentially follows thermal systematics allows the time-reversal of
this argument: $e^+e^-$ {\em production} from a thermal hadron gas 
is determined by the perturbative 'dual' strength given by Eq.~(\ref{Piem}).
This has also been verified by explicit hadronic rate 
calculations~\cite{LG98}.

At low $M$, the e.m. correlator is governed by nonperturbative
dynamics encoded in resonance formation.  
The experimental distribution is well described  by the
light neutral vector mesons $\rho$, $\omega$ and $\phi$ 
(vector dominance model).
Thus, in principle,  dilepton spectra from heavy-ion reactions directly 
reflect upon medium modifications of vector mesons.

But one can go further and establish connections to chiral symmetry
restoration. This necessitates the inclusion of the chiral partners 
of the vector mesons in the (theoretical) analysis.
Within the standard (non-/linear $\sigma$-model type) 
interpretation, and for $N_f$=2 flavors, the $SU(2)$ chiral partner
of the $\rho$ is the $a_1(1260)$, whereas the $\omega$ is a 
singlet (for $SU(3)$ the octet-$\omega_8$ and
singlet-$\omega_0$ mix into $\omega$ and $\phi$
rendering both non-invariant under chiral rotations).
This is to be contrasted with the recently 
suggested 'Vector Manifestation' of chiral symmetry, where
the longitudinal component of the $\rho$ is identified as the 
chiral partner of the pion~\cite{HY01}. 

\begin{figure}[htb]
\vspace*{-2cm}
\begin{center}
\leavevmode
\epsfysize=6.5cm
\epsfbox{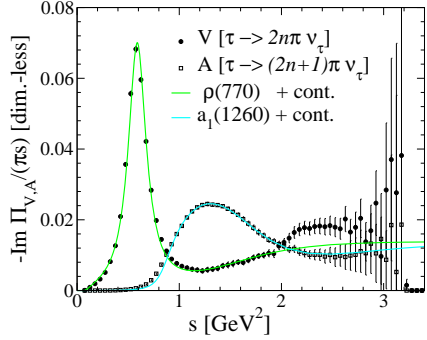}
\end{center}
\vspace*{-1cm}
\caption[]{Free isovector axial-/vector spectral functions from  
hadronic tau-decay data~\protect\cite{aleph}.}
\vspace*{-0.4cm}
\label{fig_VAfit}
\end{figure}

Spontaneous chiral symmetry breaking in the vacuum is nicely illustrated
by hadronic $\tau$ decay data in terms of the  
difference between vector and axialvector spectral functions at low $M^2$, 
cf. Fig.~\ref{fig_VAfit}.  
It can be quantified via Weinberg sum rules, \eg,  
\begin{equation}
f_\pi^2=- \int \frac{ds}{\pi s} ({\rm Im}\Pi_V -{\rm Im}\Pi_{A}) \ .
\end{equation}
Above $M^2$$\simeq$3~GeV$^2$, $\Pi_V$ and $\Pi_{A}$  degenerate 
signifying the onset of the perturbative regime. 

Finally note that the  $\rho$-meson strength in the e.m. correlator
dominates over the 
$\omega$ by about an order of magnitude (as reflected by the e.m. decay 
widths, $\Gamma_{\rho\to ee}=6.8$~keV vs. $\Gamma_{\omega\to ee}=0.6$~keV, 
or the naive quark-model prediction 
$\Gamma_{\rho\to ee}/\Gamma_{\omega\to ee}=9/1$).

\subsection{Medium Effects and Dilepton Emissivities}
A compilation of the extensive investigations on 
in-medium modifications of the e.m. correlator (dilepton production
rate) in both hadronic and quark-gluon phases is attempted
in Table~\ref{tab_med}.   
\begin{table}[h]
\vspace{-0.4cm}
\caption{Summary of medium effects in thermal dilepton production.}
\vspace{0.2cm}
\begin{tabular}{c|c|c}
          & Hadron Gas  & Quark-Gluon Plasma \\
\hline
               & \underline{in-medium $\rho$, $\omega$, $\phi$:}  &
                 \underline{perturbative QCD:}         \\
  Low          &  effective chiral Lagrangian  + VDM &
              HTL-resummed $q \bar q\to ee$~\protect\cite{BPY90} \\
  Mass      &  + finite-$T$ / -$\mu_B$ field theory~\protect\cite{RW00}    &
             \underline{non-pert. QCD:} \\
$M\lsim 1$~GeV &$D_V=[M^2-m_V^2-\Sigma_V(M,q;\mu_B,T)]^{-1}$    &
              gluon condensates ($T$$\gsim$$T_c$)~\protect\cite{LWZH99}     \\
\hline
Intermediate   &  $\pi a_1 \to l^+l^-$ annihilation~\protect\cite{LG98}  &
                      'bare' $\alpha_s$ corrections  \\
Mass           &  $\hat{=}$ chiral V-A mixing~\protect\cite{DEI90}        &
                  to $q\bar q$ annihilation~\protect\cite{AR89}  \\
$M\gsim 1$~GeV &$\Pi_{V,A}=(1-\epsilon)~\Pi_{V,A}^\circ
                       +\epsilon~\Pi_{A,V}^\circ$
                 &  order ${\cal O}(\alpha_s\frac{T^2}{M^2})$
\end{tabular}
\label{tab_med}
\end{table}
As argued above, beyond masses of 1.5~GeV the rates are reasonably well 
understood with moderate $\alpha_s$-corrections to $q\bar q$ 
annihilation in the QGP, and 
prevalent $4\pi$-type annihilations (such as $\pi a_1$ or $\pi\omega$)
in the hadron gas. The $\pi a_1$ channel, however, bears special
relevance in the non-/perturbative 'transition region' between
1 and 1.5~GeV: it can be interpreted as a lowest order in temperature
'chiral' mixing (in parallel with $\pi\rho\to a_1$)~\cite{DEI90} leading 
(for $M> 1$~GeV) to a 3-fold degeneration of $\Pi_V$, $\Pi_A$ 
and the pQCD continuum at temperatures not far from the expected $T_c$.  

\begin{figure}[htb]
\vspace*{-1.9cm}
\leavevmode
\epsfysize=6.8cm
\hspace{-0.2cm}
\epsfbox{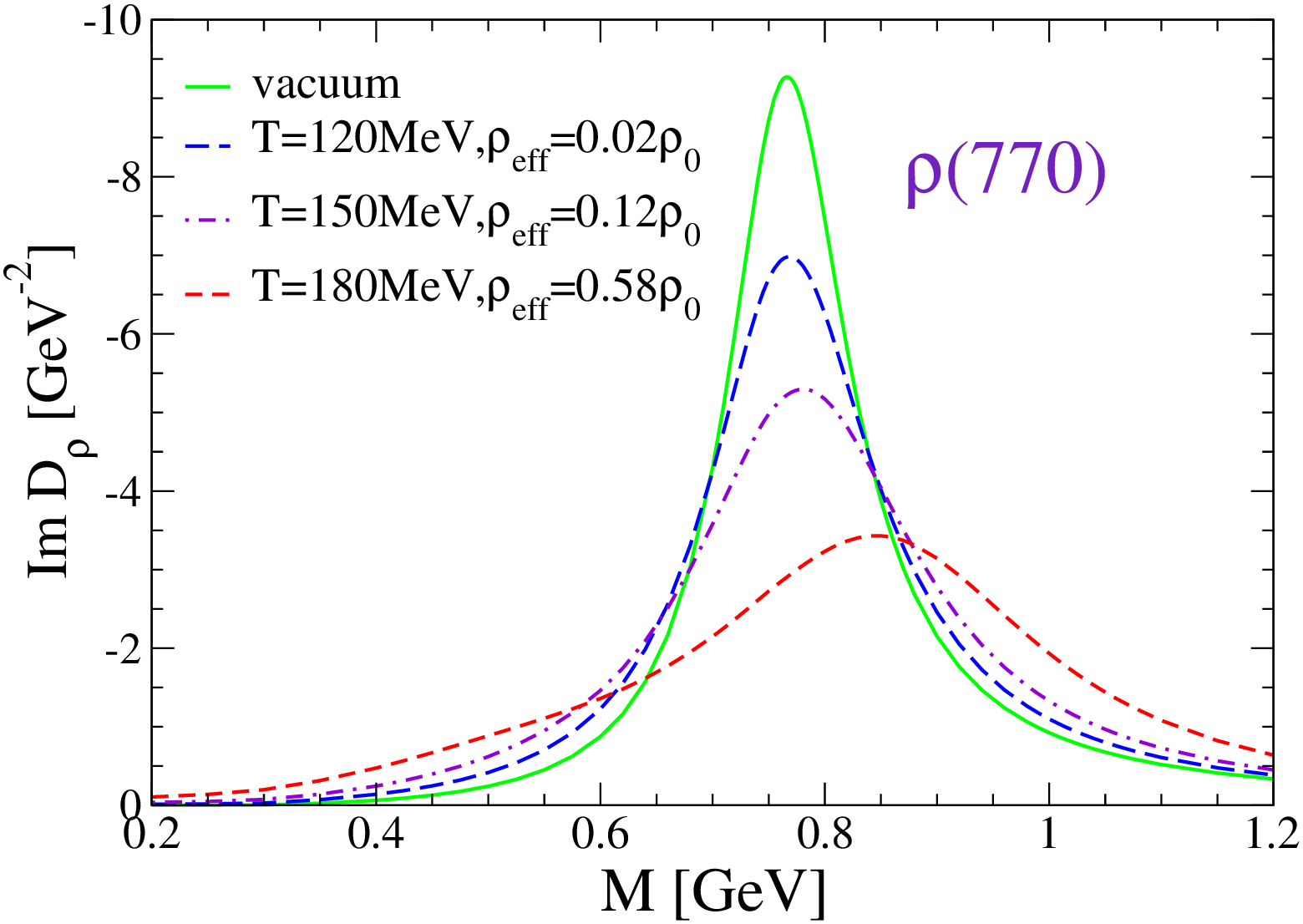}
\hspace{1cm}
\leavevmode
\epsfysize=6.8cm
\epsfbox{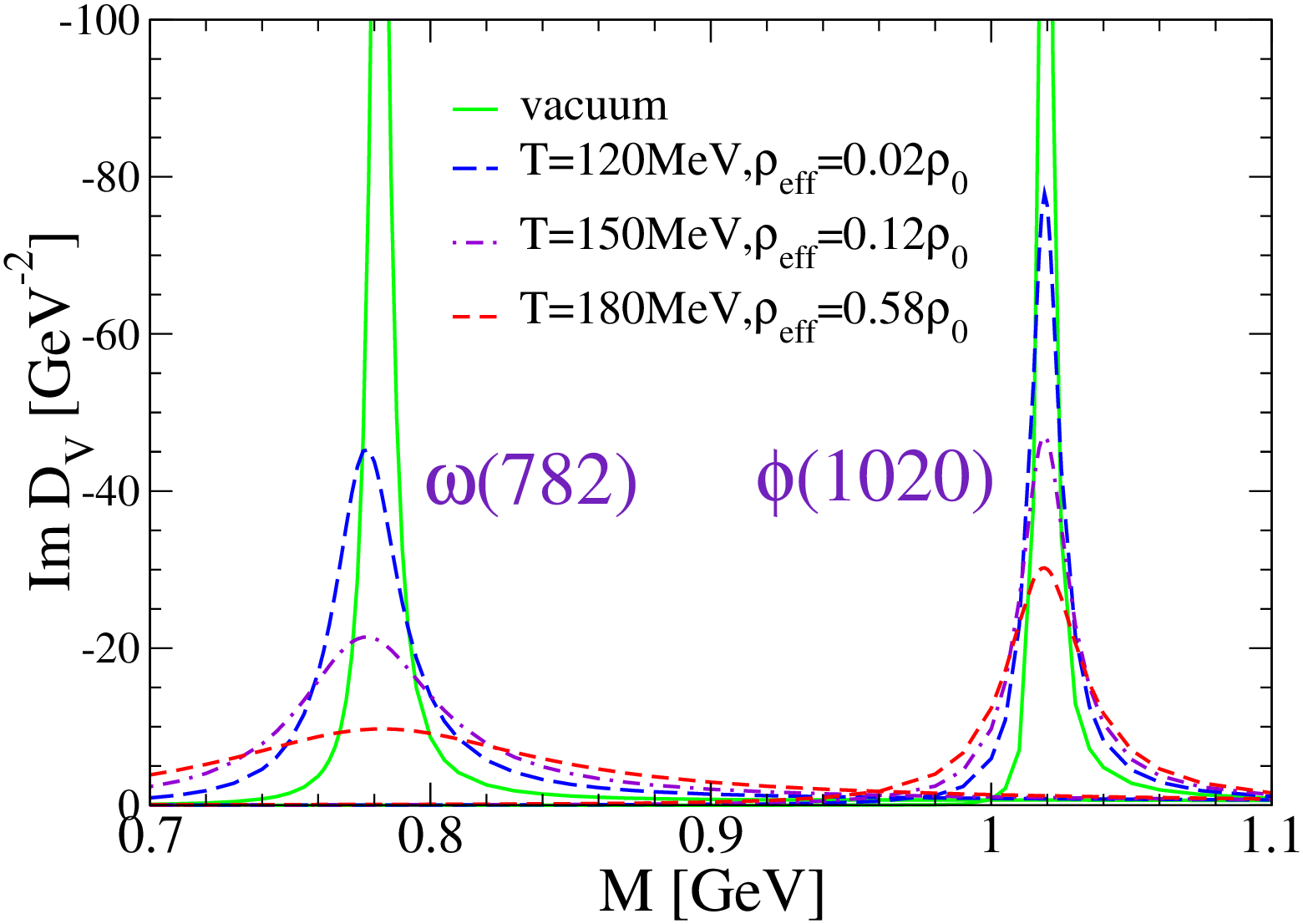}
\vspace*{-0.65cm}
\caption[]{Light vector-meson spectral functions~\protect\cite{Ra01} 
in thermal and chemical equilibrium at fixed entropy-per-baryon
characteristic for RHIC-200 conditions.}
\vspace*{-0.4cm}
\label{fig_vec1}
\end{figure}

At low mass in the hadronic sector, substantial medium effects are
predicted for $\rho$ and $\omega$ mesons (less pronounced for $\phi$ mesons
due to the OZI rule). Most approaches have reached consensus  
that baryon-induced effects dominate
over meson-induced ones at comparable densities, even on a quantitative
level~\cite{RW00,HBEK02}.   
This seems to imply that medium effects at heavy-ion colliders
(RHIC and LHC), where the net baryon densities at midrapidity are
small, are less pronounced. However, as pointed out in ref.~\cite{Ra01},
vector mesons equally interact with baryons and antibaryons, \ie,  the
relevant quantity for medium effects is the {\em sum} of baryon and 
antibaryon density. Indeed, at midrapidities at RHIC~\cite{phenix1201},
$dN_{B}/dy$ is about as large as at SPS. 
An example of vector-meson spectral functions
at small {\em net} baryon density is displayed in
Fig.~\ref{fig_vec1}. At the highest temperatures ($T=180$~MeV),
$\sim$40\% of the medium effects stem from anti-/baryons.

\begin{figure}[htb]
\vspace*{-0.6cm}
\begin{center}
\leavevmode
\epsfysize=9cm
\epsfbox{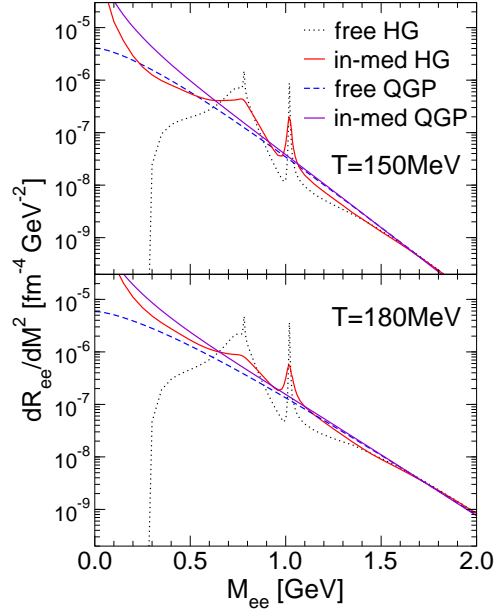}
\end{center}
\vspace*{-1cm}
\caption[]{3-momentum integrated dilepton production rates from
equilibrium matter with specific entropy characteristic for RHIC
energies.} 
\vspace*{-0.4cm}
\label{fig_rates}
\end{figure}

Pertinent 3-momentum integrated dilepton production rates
are shown in Fig.~\ref{fig_rates} and compared to  hadronic
rates without medium effects (dotted curves), as well as to  
QGP emission to lowest order ${\cal O}(\alpha_s^0)$
and  within HTL-resummed perturbation theory~\cite{BPY90}.   
One finds that even for small {\em net} baryon densities the in-medium
hadronic and QGP results become remarkably similar when approaching 
the expected $T_c$, supporting the notion of quark-hadron duality 
towards low masses in the vicinity of the phase transition~\cite{Ra99}.
The key features are resonance melting in the hadronic, as well
as soft emission characteristic for interacting many-body systems
in both the hadronic and QGP phases.

\section{Dileptons in Heavy-Ion Collisions}
\label{sec_appl}

\subsection{Thermal Production and Chemical Off-Equilibrium}
\label{sec_offeq}
Thermal dilepton production in   
heavy-ion collisions is calculated by a convolution of the rate
given in Eq.~(\ref{rate}) over the space-time history of the 
system. The natural dynamic framework are hydrodynamic~\cite{HS97,HRS99}
 (or simplified thermal fireball) simulations, being formulated in the 
same variables as the rates (temperature and baryon chemical potential).  
For invariant mass spectra, \eg, one has ($Acc$ indicates kinematic detector 
acceptance cuts)
\beq
\frac{dN^{thermal}_{l^+l^-}}{dM}=\int\limits_{\tau_0}^{\tau_{fo}}
            d\tau \ V_{FB}(\tau) \int d^3q  \ \frac{M}{q_0} \
                  \frac{dR_{l^+l^-}^{thermal}}{d^4q} \ Acc \ ,
\eeq
Clearly, a realistic space-time description must be consistent with 
hadronic observables. 

With elastic hadronic cross sections being much larger than inelastic 
ones, thermal equilibrium in the expanding hadron gas phase is 
maintained longer than the chemical one. Therefore, subsequent to 
chemical freezeout (which for SPS energies and beyond essentially
coincides with the expected phase boundary to the QGP)
the effective  numbers of non-strongly decaying particles have to 
be conserved. This can be implemented into the equation of state
via additional (effective) chemical potentials for, \eg, 
pions, kaons, etas and even antibaryons.   
These can have substantial impact on the composition and 
evolution of the hadronic system.
In particular, the build-up of large pion-chemical potentials
enhances thermal dilepton production through $\pi\pi$
annihilation~\cite{Ra99}, which is often neglected in hydrodynamic
calculations.
  
\begin{figure}[htb]
\vspace*{-1.9cm}
\leavevmode
\epsfysize=6.8cm
\hspace{-0.2cm}
\epsfbox{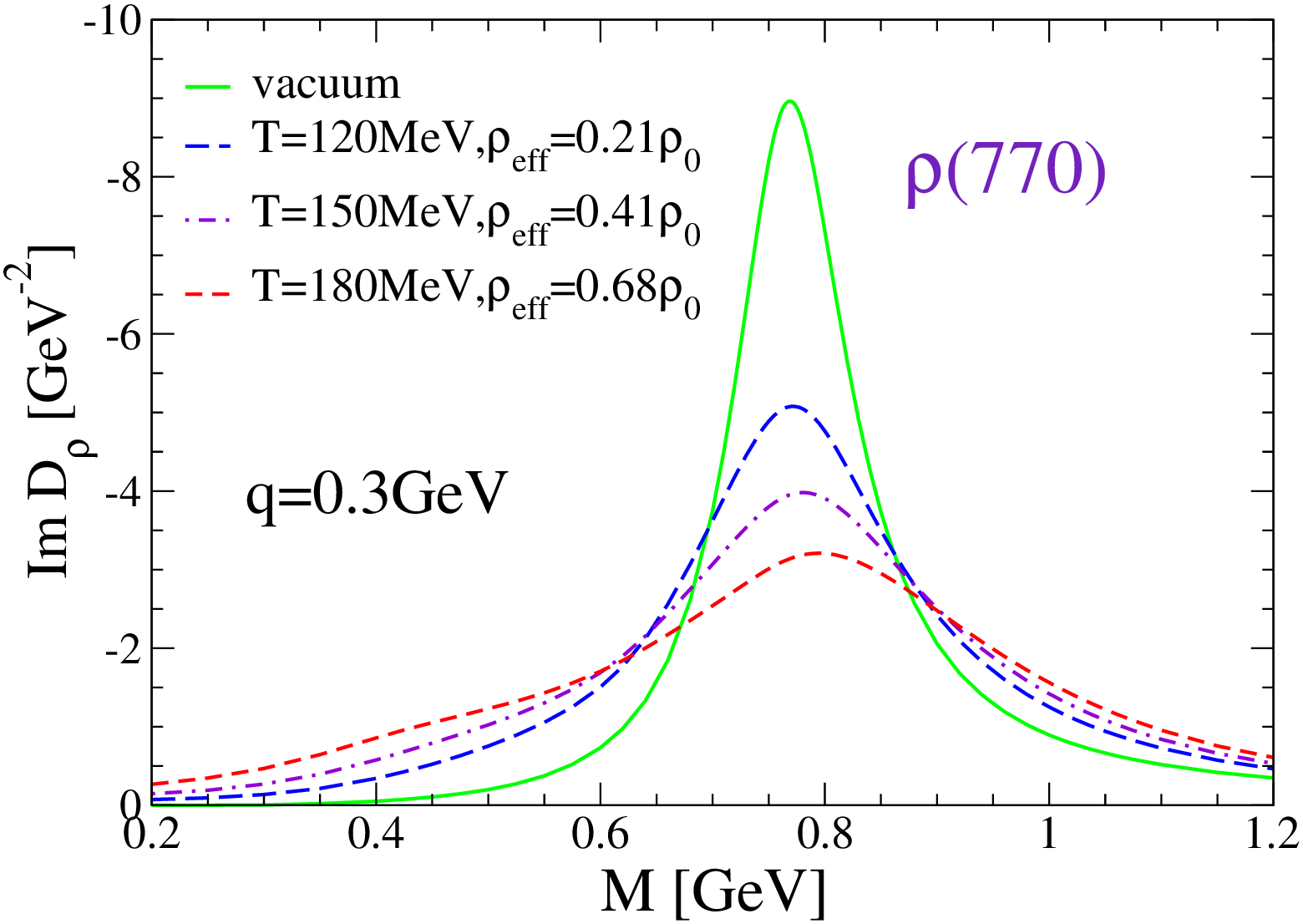}
\hspace{1cm}
\leavevmode
\epsfysize=6.8cm
\epsfbox{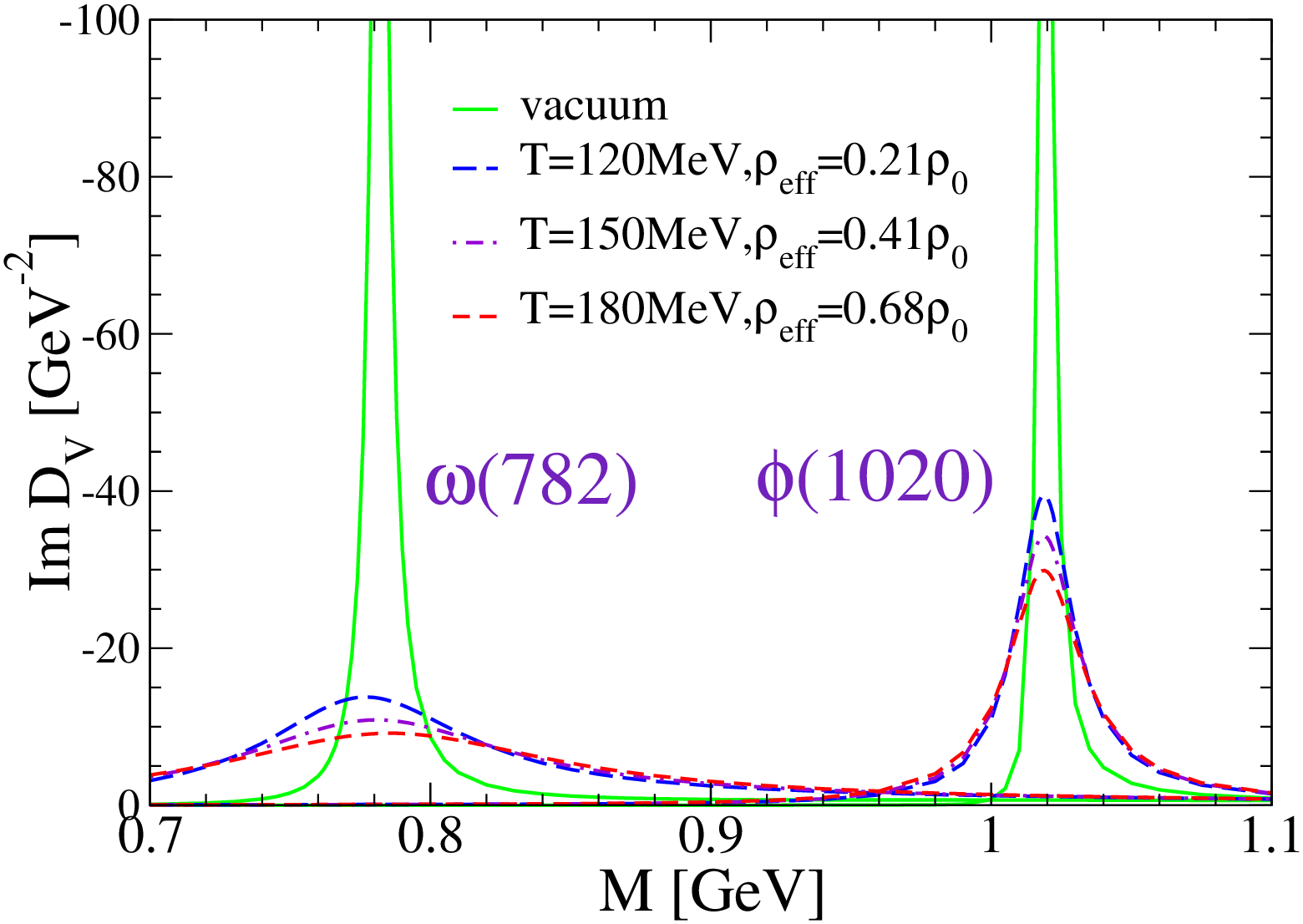}
\vspace*{-0.65cm}
\caption[]{Light vector-meson spectral function for RHIC-200
conditions including chemical-off-equilibrium effects.}
\vspace*{-0.4cm}
\label{fig_vec2}
\end{figure}

At RHIC energies, antibaryon-number conservation has appreciable
consequences for low-mass dilepton spectra.   
If no corresponding effective chemical $\mu_{\bar B}^{eff}$ 
is introduced, the anti/-baryon abundance at the later hadronic 
stages (assuming isentropic expansion) will be much reduced; at the same 
time, meson-chemical potentials seem to remain small~\cite{Ra01}. 
On the other hand, imposing the individual
anti-/baryon numbers to be constant throughout the evolution, and
starting from full chemical equilibrium at $(\mu_N,T)=(25,180)$~MeV,
leads to anti-/baryon chemical potentials of up to $\mu_N\simeq$~270~MeV
and $\mu_{\bar N}=-\mu_N+\mu_{\bar B}^{eff}\simeq$~230~MeV at a
temperature $T\simeq 130$~MeV. Due to entropy conservation one then 
also finds appreciable pion- and kaon-chemical potentials,
\eg, $\mu_\pi(T=130~{\rm MeV})\simeq$~70~MeV.
To illustrate the impact on dilepton production, 
the vector-meson spectral functions along such a trajectory
are displayed in Fig.~\ref{fig_vec2}. As compared to Fig.~\ref{fig_vec1}, 
one finds that chemical off-equilibrium reinforces medium effects at the 
later stages.

An accurate identification of the thermal dilepton component furthermore
hinges on a reliable knowledge of 'background' sources, such as e.m. 
hadron decays after freezeout (low mass), or Drell-Yan 
annihilation and correlated open charm decays (higher masses).

\subsection{Post- and Predictions at SPS Energies}
\label{sec_sps}
Numerous theoretical works have addressed the low-mass excess
observed by the NA45 collaboration (CERES). The minimal conclusion at 
this point is that strong in-medium effects on the $\rho$-meson are 
needed to fill in the mass region around $0.4$~GeV without overpredicting
the yield around the free $\rho$-$\omega$ mass (which is largely
saturated by the hadronic decay 'cocktail'). 
\begin{figure}[htb]
\vspace*{-1.5cm}
\leavevmode
\epsfysize=6.8cm
\hspace{-0.2cm}
\epsfbox{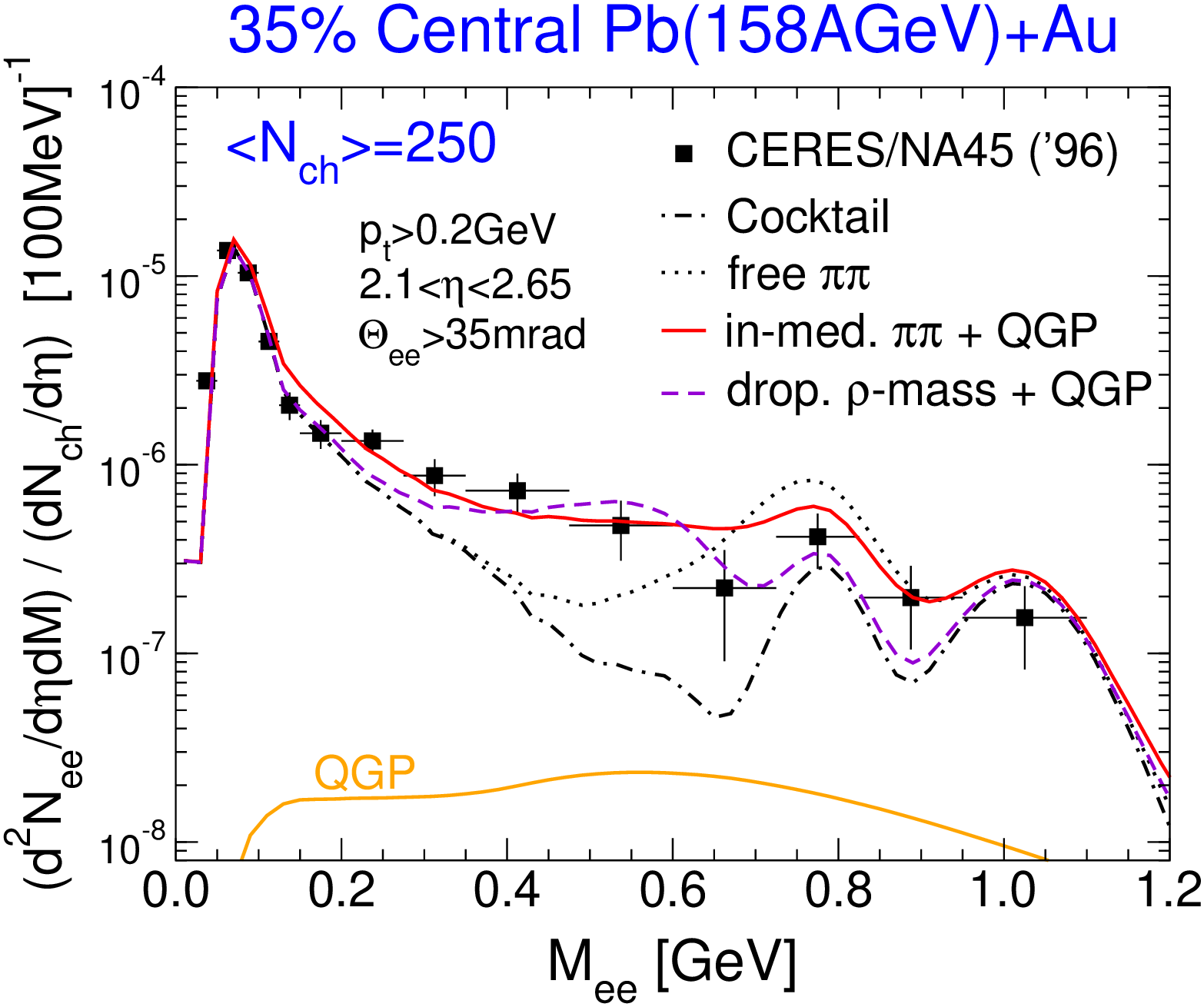}
\hspace{1cm}
\leavevmode
\epsfysize=6.8cm
\epsfbox{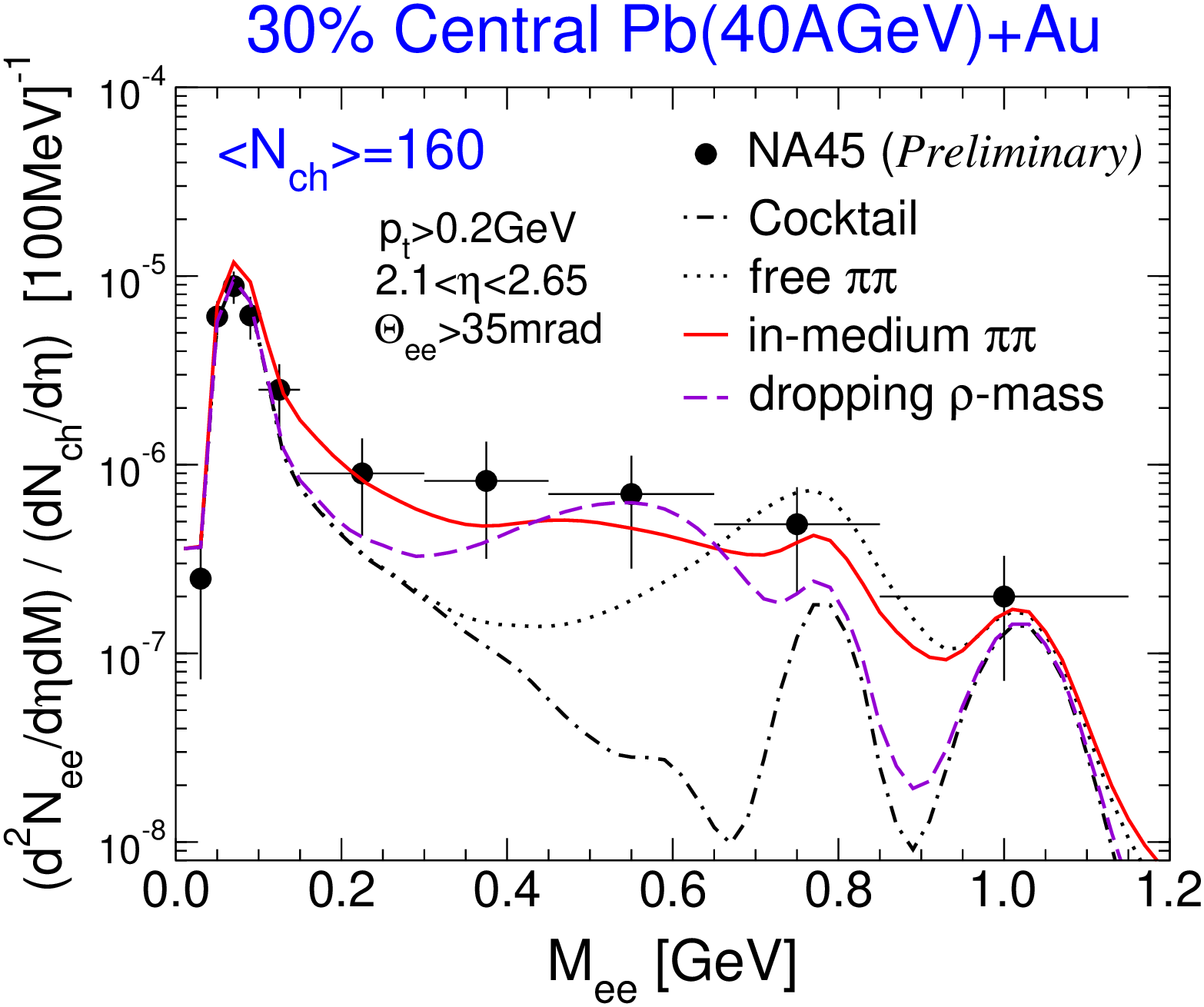}
\vspace*{-0.6cm}
\caption[]{Thermal fireball calculations~\protect\cite{RW99} plus 
hadronic decay cocktail (dashed-dotted lines) compared to CERES data at 
full~\protect\cite{na45-158} (left panel) and lower~\protect\cite{na45-40} 
(right panel) SPS energies.}
\vspace*{-0.4cm}
\label{fig_ceres}
\end{figure}
 However, within the present experimental
accuracy, it is not possible to discriminate the conjecture of
a dropping $\rho$-mass from a $\rho$-melting (due to strong
reinteractions in a hot and dense medium), as illustrated by thermal
fireball calculations in  Fig.~\ref{fig_ceres}.

The suggestion of the 'Vector Manifestation' of chiral 
symmetry~\cite{HY01} has, in fact, revoked interest in the 'dropping mass'
scenario~\cite{BR01}. On the other hand, a recent finite-temperature 
analysis to one-loop order within a chirally symmetric hadronic model with 
realistic vacuum properties including $a_1$-mesons lends further
support to broadening as the prevalent in-medium modification of 
$\rho$ and $a_1$-mesons~\cite{UBW02}. The essential difference
between these approaches lies in whether medium effects in the bare 
parameters of the underlying interaction lagrangians are to be 
incorporated or not. The ensuing  realizations of chiral symmetry 
restoration are obviously rather different. 
Nevertheless, within both scenarios stronger medium effects 
were predicted at lower SPS energies as being of baryonic origin, 
which is supported by available 40~AGeV CERES data (right panel in 
Fig.~\ref{fig_ceres}). 

\begin{figure}[htb]
\vspace*{-1.3cm}
\leavevmode
\epsfysize=6.8cm
\hspace{-0.2cm}
\epsfbox{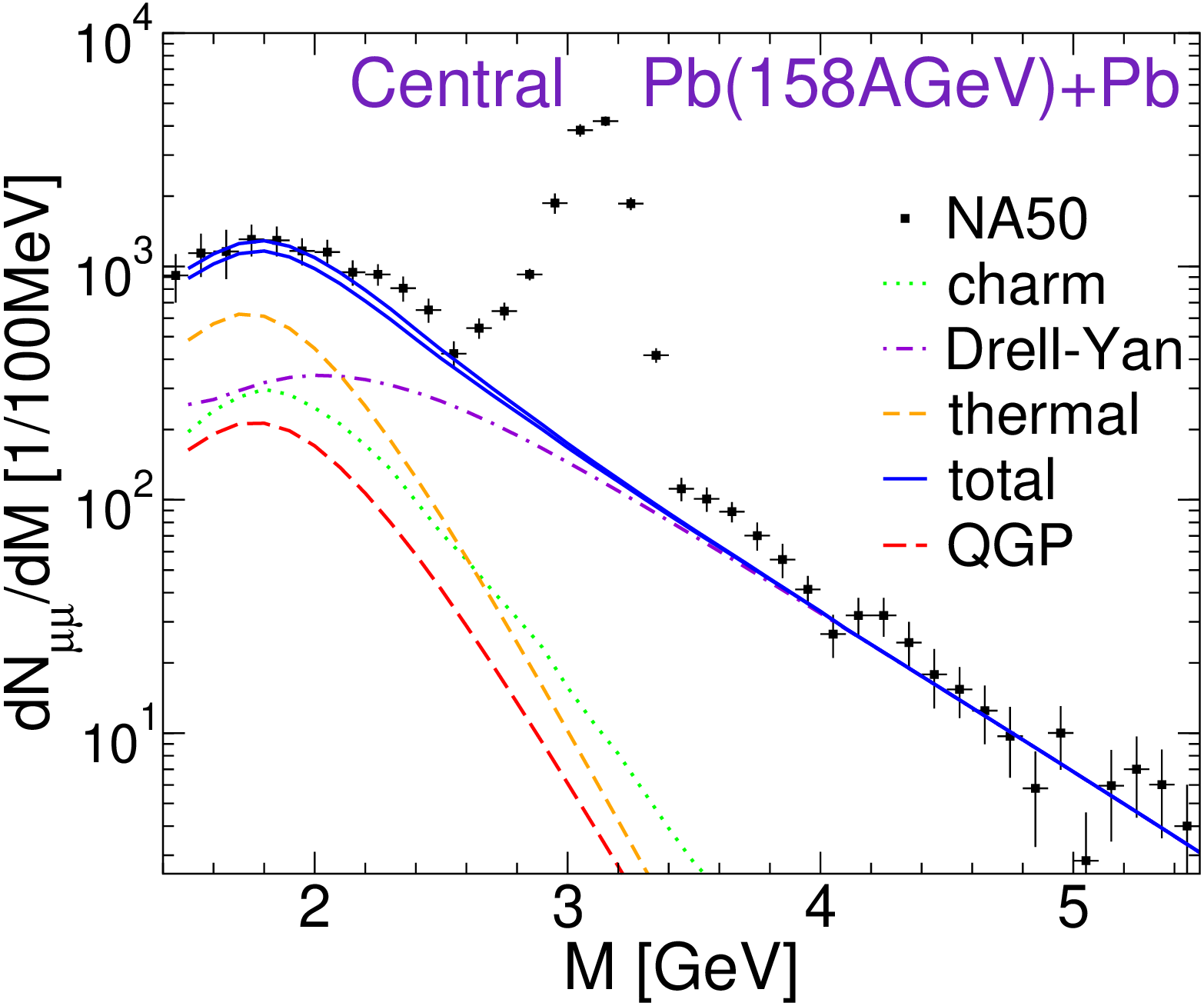}
\hspace{1cm}
\leavevmode
\epsfysize=6.8cm
\epsfbox{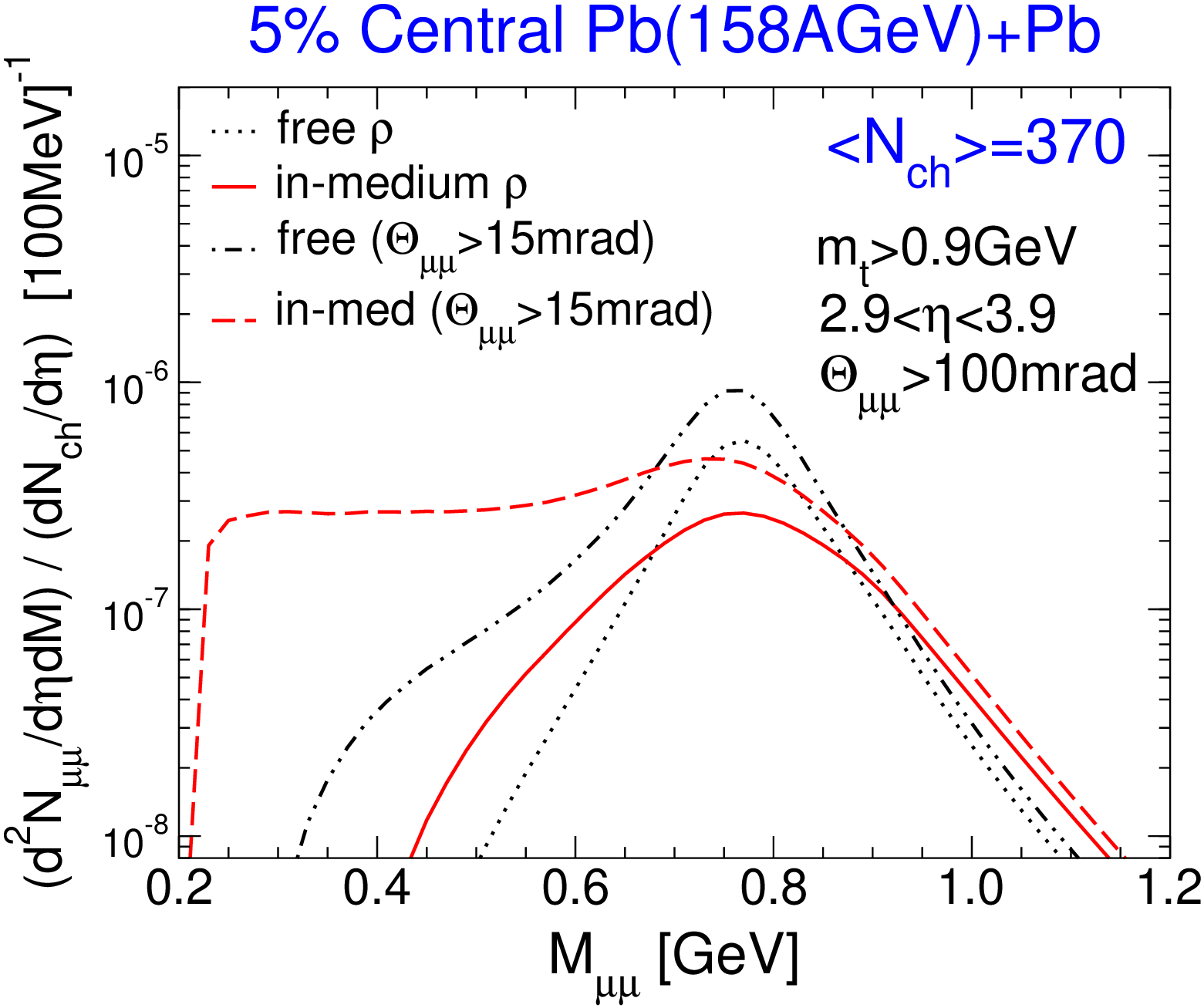}
\vspace*{-0.65cm}
\caption[]{Left panel: NA50 dimuon spectra~\protect\cite{na50-int}
compared to a sum of $pp$-extrapolated Drell-Yan and open-charm sources 
plus thermal radiation~\protect\cite{RS00}.  
Right panel: Low-mass prediction for an approximate NA60 acceptance.}
\vspace*{-0.4cm} 
\label{fig_na50}
\end{figure}
Within the same thermal evolution framework, the intermediate-mass
enhancement observed by NA50 can also be explained, cf. left panel of
Fig.~\ref{fig_na50}. At these higher
masses, where the production rates are fairly well-established,
one becomes more sensitive
to the early stages of the collision, pointing at initial
temperatures of $\sim$220~MeV or so, well inside the QGP phase.
For the future NA60 run it may also be interesting to provide a 
low mass prediction; this is shown in the right panel of
Fig.~\ref{fig_na50} for the thermal contribution. One realizes that 
with a sufficiently small
opening angle cut, a substantial low-mass enhancement could be
observed. But even with a large opening-angle cut, 
the peak-reduction of the in-medium prediction by a factor of 2 as compared
to the vacuum spectral function persists. 
The situation is very similar at 40~AGeV (not shown).

\subsection{Improved Predictions for RHIC}
\label{sec_rhic}
\begin{figure}[htb]
\vspace*{-1.8cm}
\leavevmode
\epsfysize=7cm
\hspace{-0.2cm}
\epsfbox{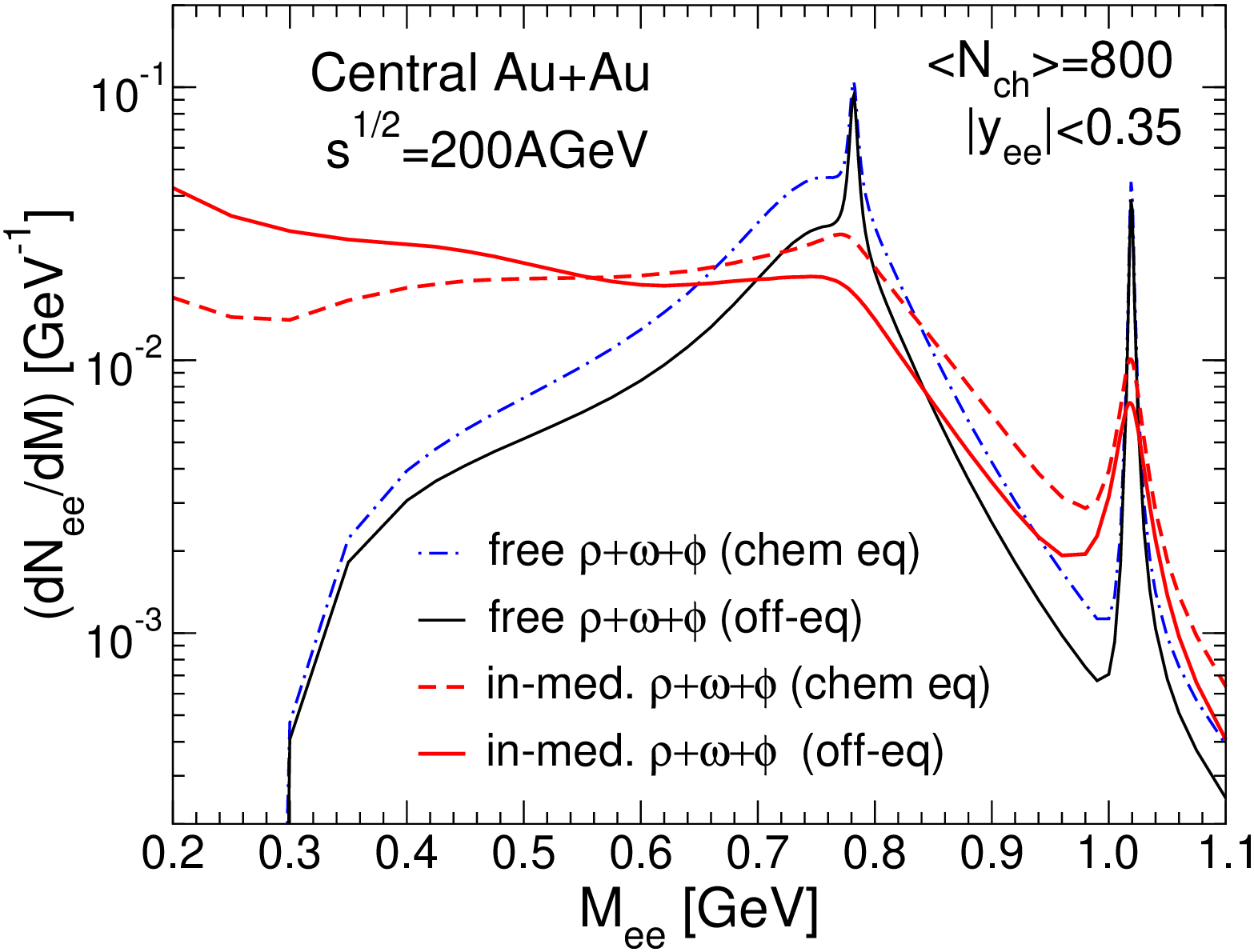}
\hspace{1cm}
\leavevmode
\epsfysize=6.8cm
\epsfbox{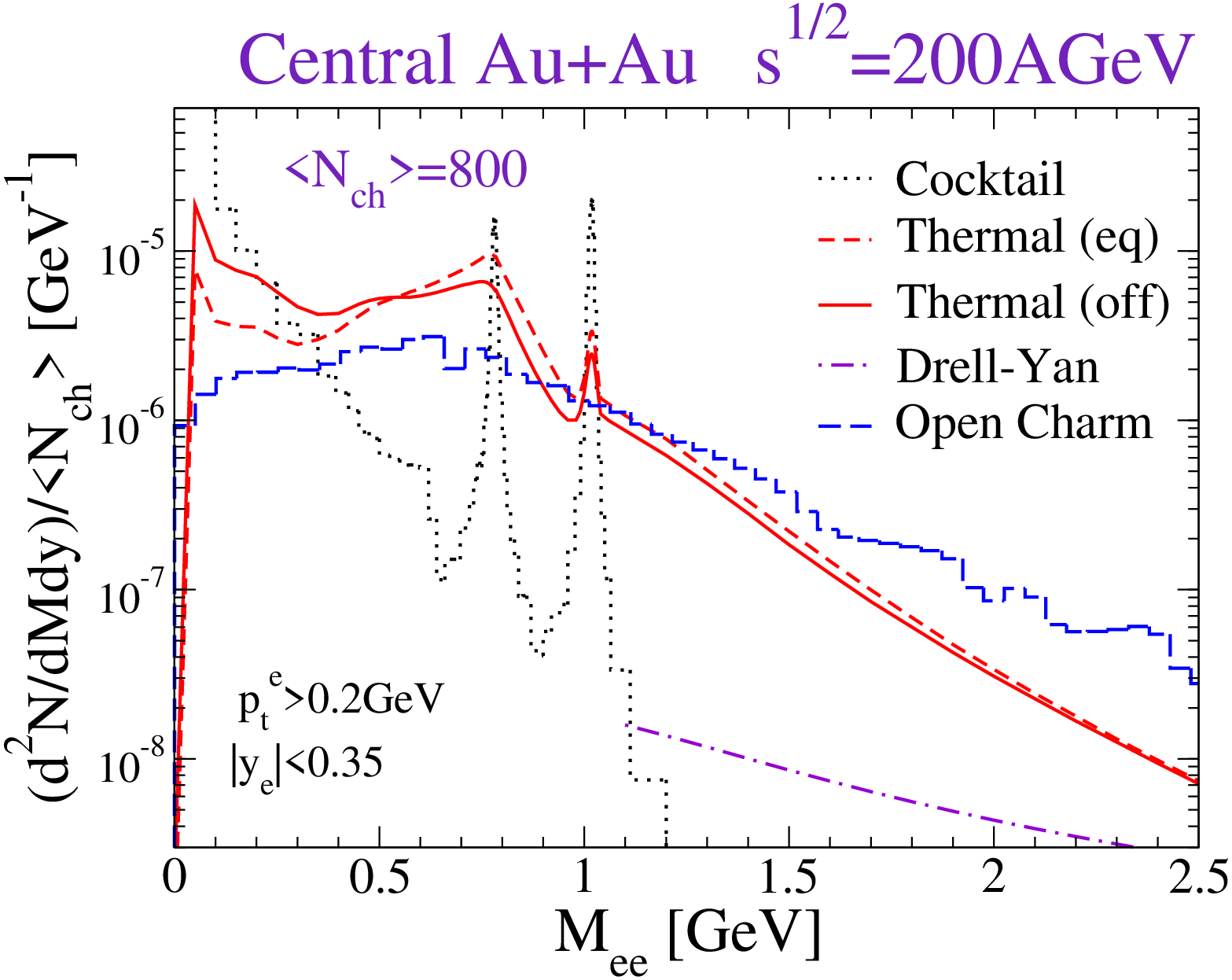}
\vspace*{-0.65cm}
\caption[]{Left panel: Hadron Gas emission at RHIC comparing the results of 
ref.~\protect\cite{Ra01} with those using an improved fireball evolution 
including chemical off-equilibrium and reduced lifetime;  
right panel: corresponding thermal spectra compared to hadronic
cocktail and open-charm decays~\protect\cite{Av01} as well as Drell-Yan
annihilation.}
\vspace*{-0.4cm}
\label{fig_200}
\end{figure}

Before RHIC started operation, the common belief was that an enhanced  
particle production entails significantly longer fireball 
lifetimes than at SPS. Hadronic observables at RHIC show that 
$dN_{ch}/dy$ at $\sqrt{s}=200$~AGeV is indeed by a factor of 2  
larger than at SPS ($\sqrt{s}=17$~AGeV), but, surprisingly, 
the HBT source radii do not change much~\cite{star-hbt}.
Although hydrodynamic simulations are currently not able to explain
this feature, it can be easily incorporated in a suitable fireball
parameterization of the expansion. 
We have employed such a phenomenologically improved description 
to assess the effect on dilepton spectra. In addition we have also 
incorporated antibaryon number conservation until thermal freezeout
as delineated in Sect.~\ref{sec_offeq}. The resulting low-mass spectra
are compared to the earlier prediction of ref.~\cite{Ra01} in
Fig.~\ref{fig_200}.
As to be expected, the reduced fireball lifetime decreases the hadron 
gas yield somewhat~\cite{Ra01}, but, more importantly, the improved 
hadro-chemistry reinforces baryonic effects in terms of resonance 
broadening and low-mass enhancement.
  
The most severe physics background for dilepton spectra at collider energies
are open-charm decays~\cite{GMRV96}; while those are subleading at SPS
energies they may become dominant at RHIC, at least at intermediate masses,
cf. right panel of Fig.~\ref{fig_200}. 

The charm production cross section at RHIC energies is rather uncertain;
first indirect evidence that it complies well with $pp$-based 
extrapolations from lower energies is provided by single-electron
spectra measured by PHENIX at $\sqrt{s}=130$~AGeV: charm decays nicely 
account for the excess over observed sources at $p_t^e\ge 1.5$~GeV. At 
the same time, the thermal contribution to the single-$e^\pm$ spectra turns
out to be negligible. This situation changes in the pair spectra:
the open-charm contribution is relatively suppressed to thermal radiation 
(especially at lower masses) by an additional semileptonic branching
ratio, their harder spectral shape and the typically large rapidity gaps
between $e^+$ and $e^-$ (within the rather narrow PHENIX acceptance).  
\begin{figure}[htb]
\vspace*{-0.4cm}
\leavevmode
\epsfysize=6.7cm
\epsfbox{dNedpt-130.epsi}
\hspace{0.2cm}
\leavevmode
\epsfysize=7.4cm
\epsfbox{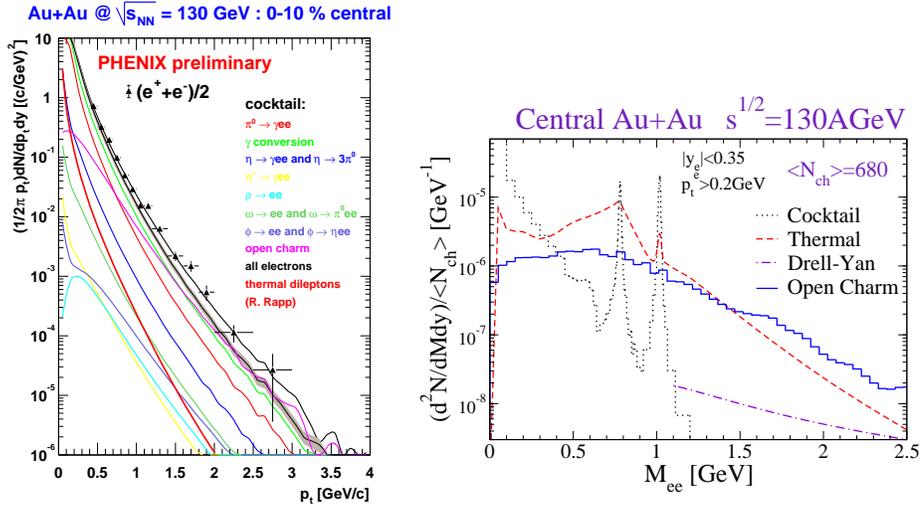}
\caption[]{Left panel: preliminary single-electron $p_t$ spectra
from PHENIX~\protect\cite{phenix0202} compared to expected sources,
as well as additional thermal and open-charm contributions. 
Right panel: corresponding $e^+e^-$ pair spectra based on the 
same ingredients as the singles.} 
\label{fig_130}
\end{figure}

\section{Conclusions}
\label{sec_sum}
Dileptons continue to be an exciting probe in heavy-ion physics.  
From the theoretical side, more analysis is required to 
strengthen connections to chiral symmetry restoration, capitalizing
further on recent progress. In particular, baryonic
effects appear to play a key role. 
From a practical point of view it is thus encouraging to note that the
(anti-) baryon driven impact on low-mass spectra remains appreciable at 
collider energies (reinforced by chemical off-equilibrium effects).
In addition, the open-charm 'problem' seems sufficiently controllable 
(at least at RHIC) to enable the identification of thermal radiation
(possibly even at intermediate masses, where it is of QGP origin).   
This raises the hope that the new generation experiments at GSI, CERN 
and RHIC will substantially advance our understanding of nonperturbative
dynamics in the transition regions across the QCD phase diagram.  

\section*{Acknowledgement}
I thank the organizers for the invitation to this pleasant
and stimulating workshop. This work is supported by the
U.S. Department of Energy under grant no.~DE-FG0288ER40388.

\vfill\eject
\end{document}